\documentclass[smallextended]{svjour3}       
\smartqed  
\usepackage{graphicx}
\begin{document}

\title{Microwave optomechanical measurement of non-metallized SiN strings at mK temperatures
}


\author{Sumit Kumar \and Yannick Kla{\ss} \and Baptiste Alperin \and Srisaran Venkatachalam \and Xin Zhou \and Eva Weig \and Eddy Collin \and Andrew Fefferman}


\institute{Sumit Kumar \and Baptiste Alperin \and Eddy Collin \and Andrew Fefferman \at
              Universit\'{e} Grenoble Alpes and Institut N\'{e}el, CNRS, Grenoble, France \\
              \email{andrew.fefferman@neel.cnrs.fr}           
           \and
           Yannick Kla{\ss} \and Eva Weig \at
              Technical
			University of Munich, Chair of Nano and Quantum Sensors, Department of
			Electrical and Computer Engineering, Theresienstr. 90/I 80333 Munich,
			Germany
	\and
	Srisaran Venkatachalam \and Xin Zhou \at
	Univ. Lille, CNRS, Centrale Lille, Univ. Polytechnique Hauts-de-France, UMR 8520-IEMN, F-59000 Lille, France
}

\date{Received: date / Accepted: date}

\maketitle

\begin{abstract}
The mechanical properties of amorphous materials (glasses) at low temperatures are dominated by effects of low energy excitations that are thought to be atomic-scale tunneling two level systems (TTLS). In nanometer-scale glass samples, the temperature dependence of the sound speed and dissipation is modified relative to that of bulk glass samples. In addition to this size effect, the usual presence of a polycrystalline metal in nanomechanical resonators leads to a further departure from the well-studied behavior of insulating bulk glass. We report a dual chip optomechanical measurement technique used to characterize non-metallized amorphous SiN strings at low temperatures. A harp consisting of SiN strings of width 350 nm and lengths 40 to 80 $\mu$m is coupled to an Al superconducting microwave cavity on a separate chip. The strings are driven dielectrically and their motion is detected via its modulation of the microwave resonance frequency.
\end{abstract}

\section{Introduction}
\label{intro}
At temperatures below a few kelvin, the thermal, mechanical and dielectric
properties of glass are dominated by low energy excitations (LEE) that are
not present in crystals. The first sign of these LEE was the discovery that
the heat capacity of glass below 1 kelvin is much greater than that of
crystals and has a nearly linear temperature dependence, compared with the
cubic temperature dependence observed in insulating crystals \cite{Zeller71}.
According to the tunneling model the LEE are two level systems (TLS) formed by
atoms tunneling between nearby equilibria in the disordered lattice of the
glass \cite{Anderson72,Phillips72}. Under the assumption of a broad distribution of TLS energy splittings and negligible interactions between the TLS, the predictions of the tunneling model are in good agreement with measurements of the mechanical properties of bulk insulating amorphous solids, except at the lowest temperatures \cite{Fefferman08}.

Nanomechanical resonators have several important applications. For example, they can serve as sensitive detectors \cite{Chaste12} or can be used to test the limits of quantum theory at macroscopic scales \cite{Ockeloen18}. These applications require a high quality factor. In many cases, clamping losses are negligible, so that LEE dominate the dissipation of nanomechanical resonators in the low temperature limit. It is therefore essential to understand the behavior of the LEE in nanomechanical systems to find ways to improve their performance. Glasss nanomechanical resonators also allow microscopic tests of the tunneling model since it is possible to probe individual TLS or small groups of them \cite{Remus09,Ramos13}.

Recently, Maillet \emph{et al}. demonstrated agreement between measurements of nanometer-scale, high stress SiN strings covered by aluminum and predictions of the tunneling model \cite{Maillet20}. The model had to be modified relative to the standard one since the dominant thermal phonon wavelength was greater than the lateral dimensions of the string. Furthermore, it was necessary to account for the effect of the high stress on the mechanical properties. When the superconductivity of the aluminum layer was destroyed by applying a magnetic field, the dissipation became much higher, indicating that at least some of the TLS were subject to electron-driven relaxation. These TLS must have been inside the aluminum layer, inside its surface oxide, or at the interface between the aluminum and SiN. In the superconducting state, a departure from the fit was also observed close to T$_c$, probably due to quasiparticles contributing to TLS relaxation.

Because of the complexity associated with the aluminum layer, it is desirable to make complimentary measurements of bare SiN strings at mK temperatures. Previous works have reported dielectric coupling to bare SiN strings, enabling drive and microwave optomechanical detection at room temperature \cite{Unterreithmeier09}. Furthermore, piezoelectric drive combined with microwave optomechanical detection of bare SiN strings has been demonstrated at mK temperatures \cite{Pernpeintner14}. However, in the latter work, the minimum temperature was 500 mK, probably due to the heat dissipation associated with the piezoelectric drive.

In this work, we use dielectric coupling to drive and detect the motion of bare SiN strings at mK temperatures. We expect the technique to allow measurements down to the low mK range. Such measurements of bare SiN strings will compliment those of the bilayer structure studied by Maillet \emph{et al}.

\section{Experimental Details}
\label{sec:1}
\begin{figure}
  \includegraphics[width=\textwidth]{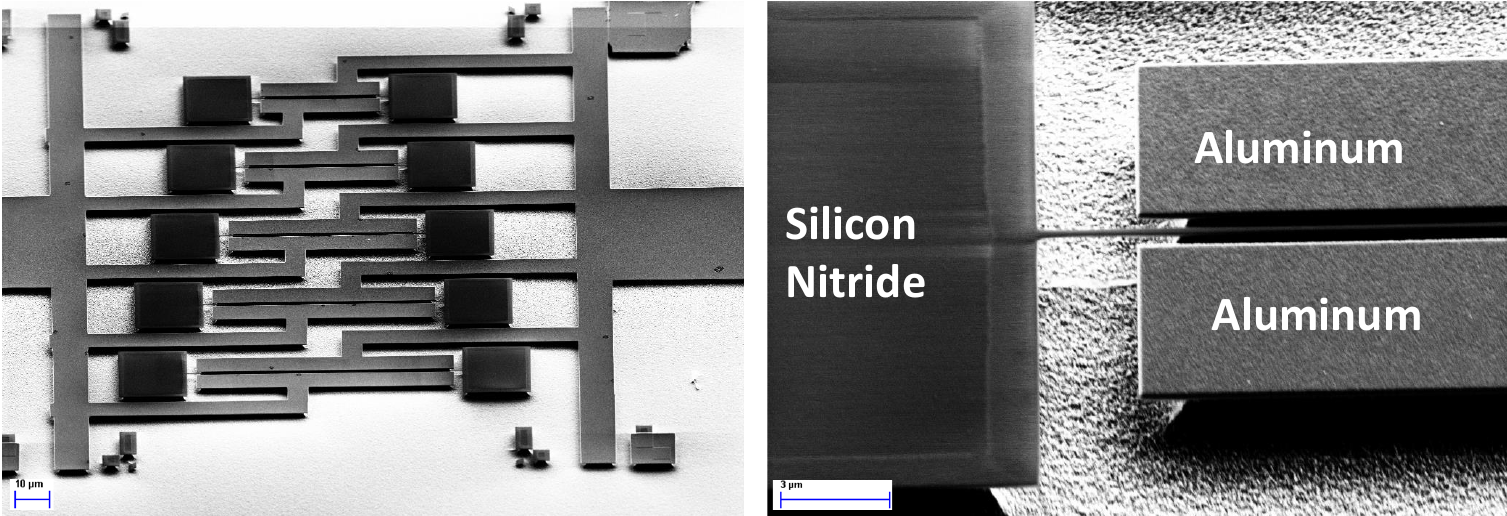}
\caption{(left) Each SiN string in the harp was between a pair of Al electrodes (upper and lower). The lower electrodes were all connected to one bonding pad, which was in turn connected to the center conductor of the microwave cavity on a separate chip. The upper electrodes were all connected to a second bonding pad, which was in turn connected to the ground plane of the microwave cavity. (right) Enlarged image of one of the SiN strings shown at left as well as its Al electrodes.}
\label{fig:1}
\end{figure}

\begin{figure}
  \includegraphics[width=\textwidth]{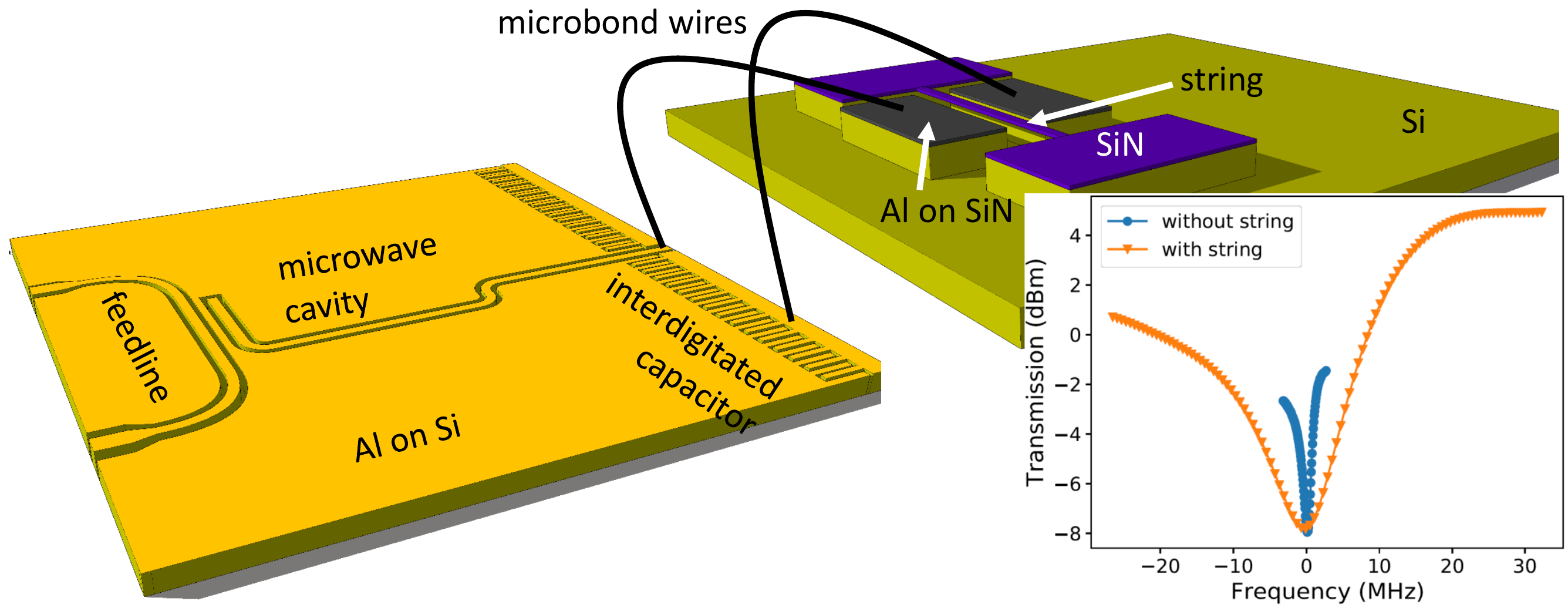}
\caption{Image of the microwave cavity chip and a schematic indicating the method used to connect it to the string electrodes. (See Fig. \ref{fig:1} for an image of the actual mechanical device.) (inset) Microwave cavity resonances before (narrow line) and after (broad line) connecting the cavity to the strings chip. Offsets of 6.1853 GHz and 6.6077 GHz have been subtracted from the resonances measured before and after connecting the strings chip, respectively.}
\label{fig:2}       
\end{figure}

A harp consisting of SiN strings of width 350 nm, thickness $100$ nm and different lengths (81 $\mu$m, 71 $\mu$m, 61 $\mu$m, 51 $\mu$m, 40 $\mu$m) (Fig. \ref{fig:1}) was fabricated using a procedure related to that of \cite{Faust12} but with certain changes. In particular, the electrodes were made of a superconducting metal, aluminum, instead of gold. Furthermore, the starting chip consisted of a high-stress SiN layer on silicon, without the sacrificial SiO$_2$ layer used in \cite{Faust12}, necessitating a different procedure for releasing the strings. In the present work, e-beam lithography and standard lift-off processes were used to define the electrodes and a chromium etch mask protecting the strings. The gap between each electrode and the string was about 350 nm. A SF$_6$+Ar reactive ion etch removed the SiN that was not covered by metal. The chromium was then selectively removed and the strings were released from the Si substrate using an XeF$_2$ etch. It was necessary to make the Al electrodes sufficiently wide so that they did not collapse during the XeF$_2$ etch. The electrodes were connected to the microwave cavity as described in Figs. \ref{fig:1} and \ref{fig:2}.

\begin{figure}
  \includegraphics[width=\textwidth]{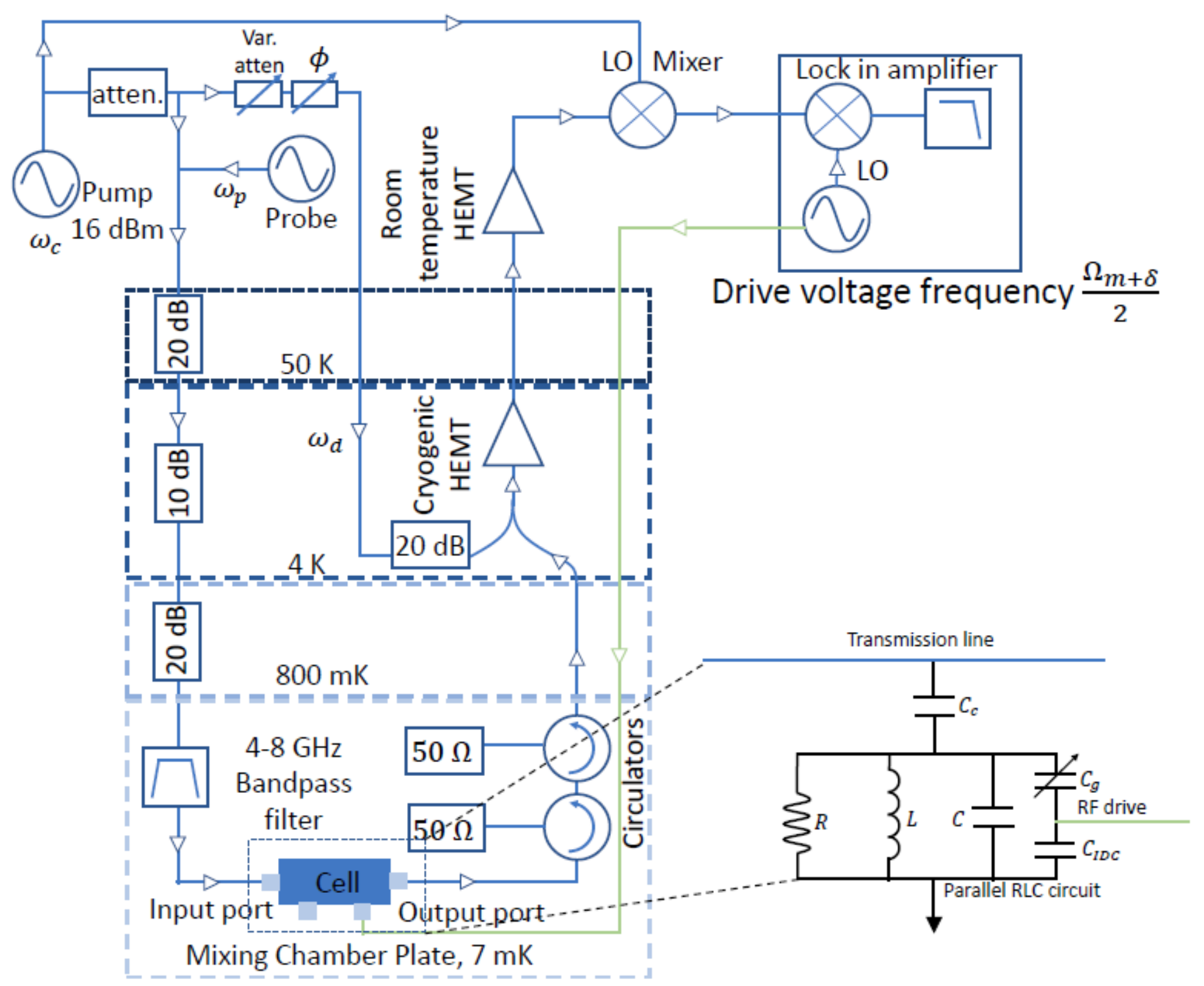}
\caption{The microwave circuit used to characterize the microwave resonator and detect the motion of the nanomechanical strings. A simple lumped element model of the on-chip circuit is shown. The parallel RLC circuit resonantor represents the microwave cavity.  The capacitance $C_g$ between the electrodes on the strings chip is modulated due to the motion of the strings. $C_{IDC}$ represents the interdigitated capacitor and $C_c$ represents the coupling between the microwave cavity and its feedline.}
\label{fig:3}       
\end{figure}

The microwave cavity was fabricated on a separate, intrinsic Si chip with resistivity greater than 10 k$\mathrm{\Omega}$cm. Laser lithography followed by deposition of 120 nm of Al and liftoff was used to fabricate the cavity. It was a $\lambda/2$ coplanar waveguide design that allowed capacitive coupling to the feedline at one end and to the motion of the mechanical resonator at the other end (Fig. \ref{fig:2}). The unique feature of the design was the presence of an interdigitated capacitor (IDC) near one edge of the chip. It functioned as a microwave bypass, taking the place of the single layer capacitor employed in \cite{Rieger12}. The motion of the strings therefore effectively modulated the capacitance of the microwave cavity, yielding optomechanical coupling. At the same time, the capacitance of the IDC was sufficiently small so that we could electrically drive the motion of the strings, which vibrated at MHz frequencies. A lumped element model of the on-chip circuit is shown in Fig. \ref{fig:3}. According to this simple model, the change in the capacitance of the microwave cavity due to the motion of the string $dC/dx$ is attenuated by a factor $1/(1+C_g/C_{IDC})^2$. Since we estimated $C_g<1$ pF and $C_{IDC}\approx60$ pF, we did not expect significant attenuation of $dC/dx$.

The cryogenic microwave circuit shown in Fig. \ref{fig:3} was used to drive and detect the nanomechanical motion. The motion of the nanomechanical string caused up- and down-conversion of the pump tone creating sidebands. In this work we pumped at the cavity resonance $\omega_c$. The transmitted fraction of the pump tone was largely cancelled at 4 kelvin using an opposition line in order to avoid saturation of the high-electron-mobility transistor (HEMT) amplifier. The remaining signal was downconverted by mixing with a signal at the pump frequency, so that the sidebands could be detected by a lockin amplifier. An oscillator inside the lockin was used to generate the mechanical drive voltage at frequency $(\Omega_m+\delta)/2$, where $\Omega_m$ is the resonance frequency of the string of interest and $\delta$ was swept in order to probe the mechanical resonance. Since the driving force acting on the strings was proportional to the square of the driving voltage, the oscillating component of the driving force was at frequency $\Omega_m+\delta$. The same oscillator in the lockin was used to generate the reference that was fed into the internal mixer, allowing us to detect the mechanical sidebands of the pump tone.

\section{Results and Discussion}
 A probe tone (Fig. \ref{fig:3}) was used to measure the transmission of the microwave cavity. Before connecting the microwave cavity to the chip containing the strings we measured a cavity linewidth of approximately 1 MHz. After coupling the cavities to the strings the cavity linewidth increased significantly (Fig. \ref{fig:2}). In order to learn about the origin of this effect, we connected the cavity to a different chip in which a nominally identical harp was closer to the edge, allowing us to use shorter microbond wires. However, the broadening of the microwave resonance did not improve. It is clear that the broadening is due to some aspect of the design of the strings chip because we did not observe broadening of the microwave resonance after connecting the cavity to a chip with a single metallized SiN string.

\begin{figure}
  \includegraphics[width=\textwidth]{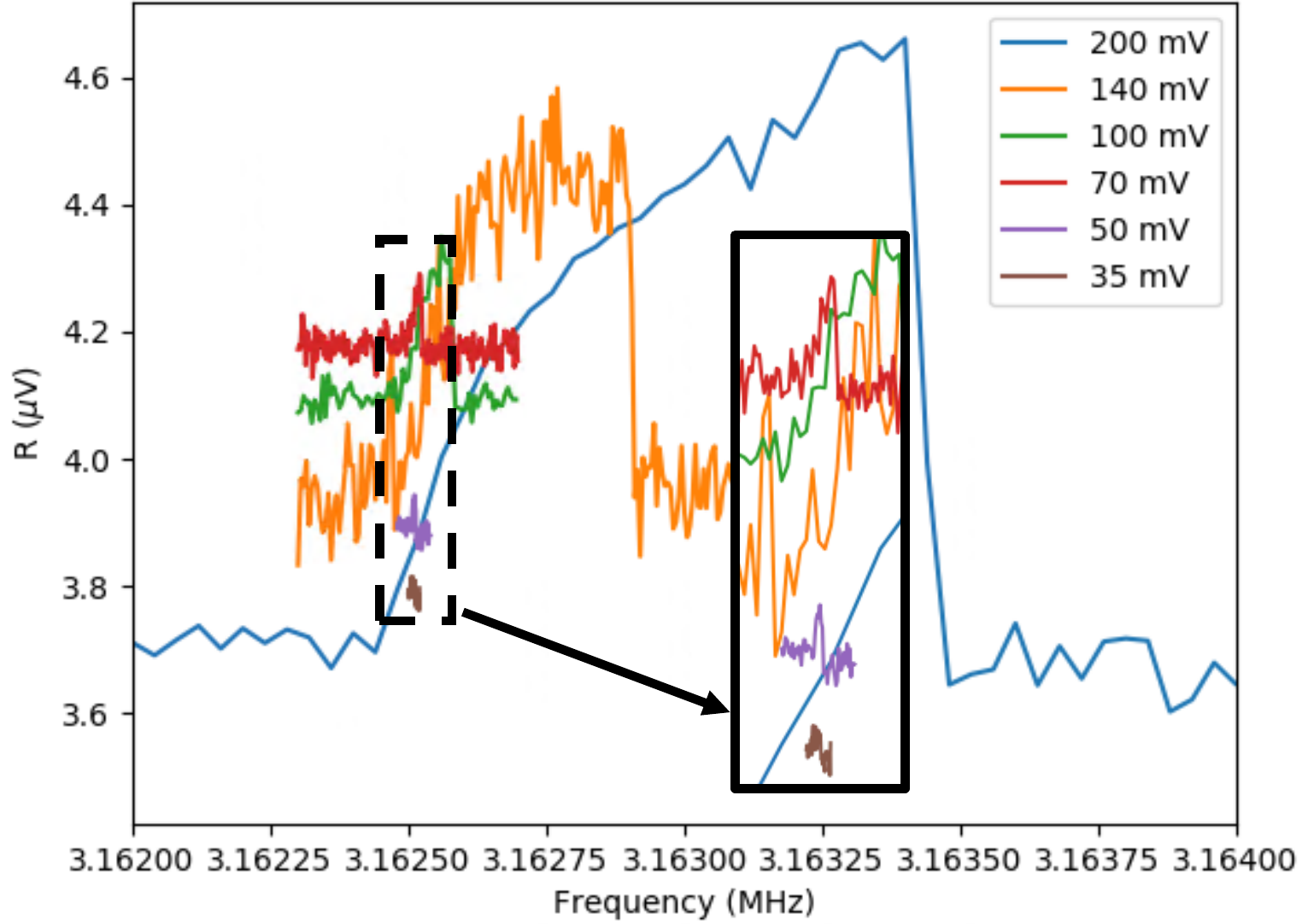}
\caption{Mechanical response of the 81 $\mu$m long string at different drive levels. The drive frequency was swept up at each drive level. The mixing chamber temperature was 284 mK. The spectrum measured at 35 mV driving voltage was shifted vertically by 0.7 $\mu$V for clarity.}
\label{fig:4}       
\end{figure}

\begin{figure}
  \includegraphics[width=\textwidth]{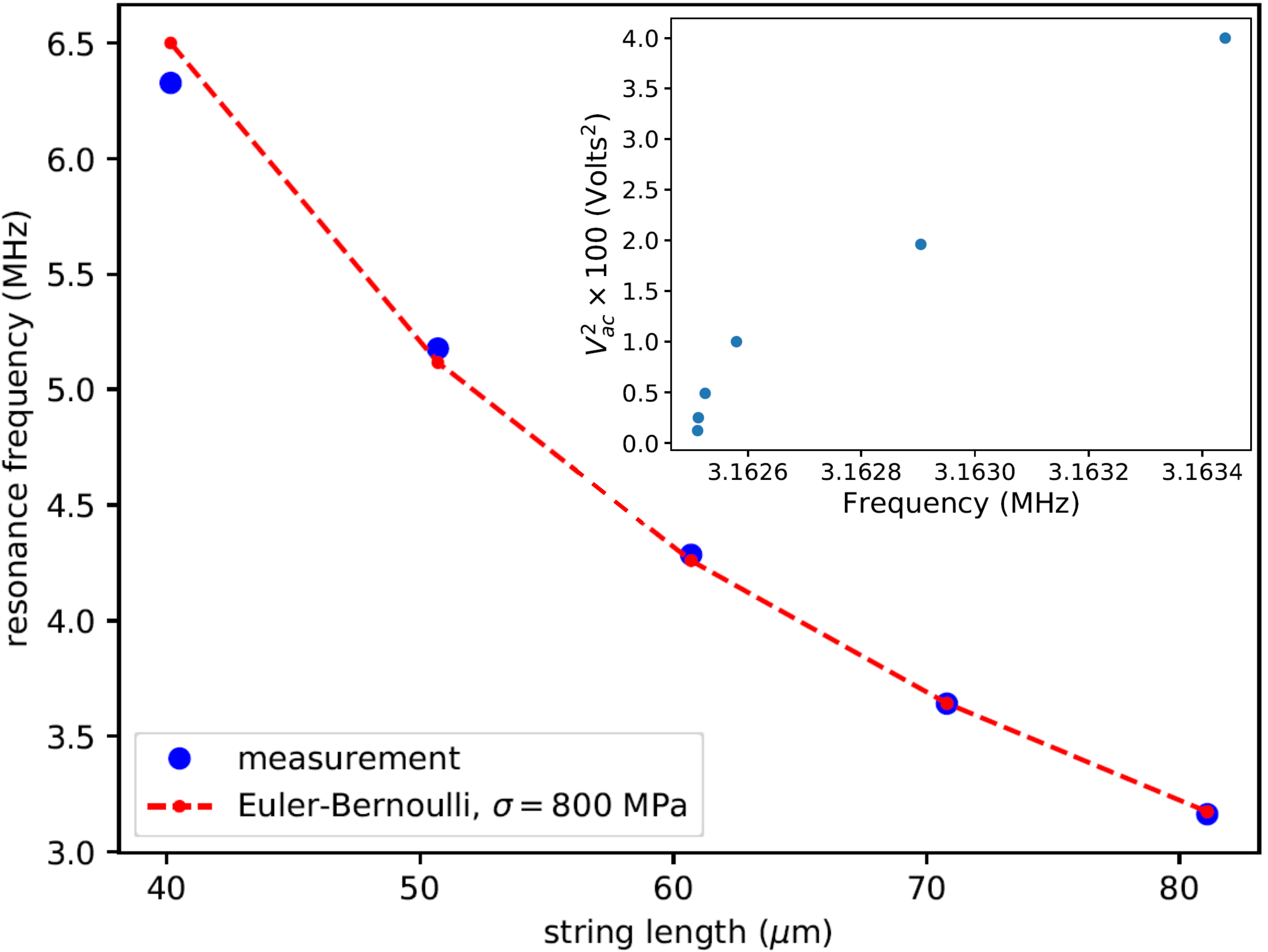}
\caption{(inset) Jump frequency of the non-linear mechanical response of the 81 $\mu$m long string as a function of $V_{ac}^2/2$, which is proportional to the driving force. (main panel) Length dependence of the resonance frequencies of the strings based on measurements like those shown in the inset. Also shown are the resonance frequencies predicted by Euler-Bernoulli theory assuming a stress in the strings of 800 MPa.}
\label{fig:5}       
\end{figure}

In order to detect the motion of the strings we used the highest possible microwave pump power of 16 dBm referenced to the output of the generator. The corresponding power into the cell was about 25 nW. Higher pump powers destabilized the opposition line, causing the HEMT to saturate, and increased the noise level. The response of the 81 $\mu$m long string measured as the driving force was swept through the resonance is shown in Fig. \ref{fig:4}. The responses at different driving forces are shown. Even at the lowest force at which we could resolve the mechanical response it was in the non-linear regime. Furthermore, we found that the background level off the mechanical resonance depended on the driving voltage. We did not observe that effect for the shorter strings, which vibrated at higher frequencies. The inset of Fig. \ref{fig:5} shows the dependence of the jump frequency, at which the non-linear mechanical resonator switched from the upper branch to the lower one, on the driving voltage. Although we cannot access the linear regime, we can use this plot to estimate the resonance frequency in the limit of low drive.

The main panel of Fig. \ref{fig:5} shows how the string resonance frequency, estimated using the technique explained above, depends on the string length. We obtain good agreement between the measured resonance frequencies and the predictions of Euler-Bernoulli theory \cite{Bokaian90} for the out-of-plane mode, with the stress in the string $\sigma=800$ MPa taken as a fit parameter. The other input parameters in the frequency calculation are the string length, the string thickness in the direction of motion, the Young's modulus of the SiN (260 GPa) and its mass density (3100 kg/m$^3$).

\section{Conclusion}
We have demonstrated the feasibility of using dielectric coupling for both drive and detection of non-metallized nanomechanical strings at mK temperatures. In order to access the linear mechanical response, we need to increase the signal to noise ratio. This can be achieved by increasing the optomechanical coupling strength. That in turn can be accomplished by decreasing the gap between the strings and the electrodes; increasing the quality factor of the microwave cavity by identifying the origin of the dissipation that occurs after connecting the strings chip to the microwave cavity; or increasing the impedance of the microwave cavity. We are currently working in these areas with the goal of characterizing the behavior of low energy excitations in purely amorphous nanomechanical resonators.

Data will be made available upon reasonable request.

We acknowledge support from the ERC StG grant UNIGLASS No. 714692, ERC CoG grant ULT-NEMS No. 647917, STaRS-MOC project No. 181386 from
Region Hauts-de-France and project No. 201050 from ISITE-MOST. The research leading to these results has received funding from the European Union's Horizon 2020 Research and Innovation Programme, under Grant Agreement no 824109. This work was partly supported by the French Renatech network, including project P-18-02468.

\bibliographystyle{spphys}       

\end{document}